\documentclass[letterpaper]{article} 
\usepackage{aaai2026}  
\usepackage{times}  
\usepackage{helvet}  
\usepackage{courier}  
\usepackage[hyphens]{url}  
\usepackage{graphicx} 
\usepackage{natbib}  
\usepackage{caption} 
\usepackage{algorithm}
\usepackage{algorithmic}

\usepackage{newfloat}
\usepackage{listings}
\DeclareCaptionStyle{ruled}{labelfont=normalfont,labelsep=colon,strut=off} 
\floatstyle{ruled}
\newfloat{listing}{tb}{lst}{}
\floatname{listing}{Listing}
\usepackage{pifont}
\usepackage{threeparttable}
\usepackage{makecell}
\usepackage{multicol}
\usepackage{multirow}
\usepackage{booktabs}
\usepackage{makecell}
\usepackage{adjustbox}
\usepackage{bm}
\usepackage{amsmath}      
\usepackage{amssymb}      

\title{Equivariant Atomic and Lattice Modeling Using Geometric Deep Learning for Crystal Structure Optimization}
\author{
    Ziduo Yang\textsuperscript{\rm 1}, Yi-Ming Zhao\textsuperscript{\rm 2}, Xian Wang\textsuperscript{\rm 2}, Wei Zhuo\textsuperscript{\rm 3}, Xiaoqing Liu\textsuperscript{\rm 2}, Lei Shen\textsuperscript{\rm 2,4}\thanks{Corresponding author.}
}
\affiliations{
    \textsuperscript{\rm 1}College of Information Science and Technology, Jinan University, Guangzhou, 510632, China \\
    \textsuperscript{\rm 2}Department of Mechanical Engineering, National University of Singapore, 9 Engineering Drive 1, 117575, Singapore \\
    \textsuperscript{\rm 3}College of Computing and Data Science, Nanyang Technological University, 50 Nanyang Avenue, 639798, Singapore
    \textsuperscript{\rm 4}National University of Singapore (Chongqing) Research Institute, Chongqing, 401123, China
}

\usepackage{bibentry}

\begin{document}

\maketitle

\begin{abstract}
Structure optimization, which yields the relaxed structure (minimum‑energy state), is essential for reliable materials property calculations, yet traditional \textit{ab initio} approaches such as density‑functional theory (DFT) are computationally intensive. Machine learning (ML) has emerged to alleviate this bottleneck but suffers from two major limitations: (i) existing models operate mainly on atoms, leaving lattice vectors implicit despite their critical role in structural optimization; and (ii) they often rely on multi-stage, non-end-to-end workflows that are prone to error accumulation. Here, we present E$^{3}$Relax—an end-to-end equivariant graph neural network that maps an unrelaxed crystal directly to its relaxed structure. E$^{3}$Relax promotes both atoms and lattice vectors to graph nodes endowed with dual scalar–vector features, enabling unified and symmetry‑preserving modeling of atomic displacements and lattice deformations. A layer‑wise supervision strategy forces every network depth to make a physically meaningful refinement, mimicking the incremental convergence of DFT while preserving a fully end‑to‑end pipeline. We evaluate E$^{3}$Relax on four benchmark datasets and demonstrate that it achieves remarkable accuracy and efficiency. Through DFT validations, we show that the structures predicted by E$^{3}$Relax are energetically favorable, making them suitable as high-quality initial configurations to accelerate DFT calculations. Our code and data are available at \url{https://github.com/Shen-Group/E3Relax}.
\end{abstract}

\section{Introduction}

\begin{figure}[!tb]
  \centering
  \includegraphics[width=0.9\columnwidth]{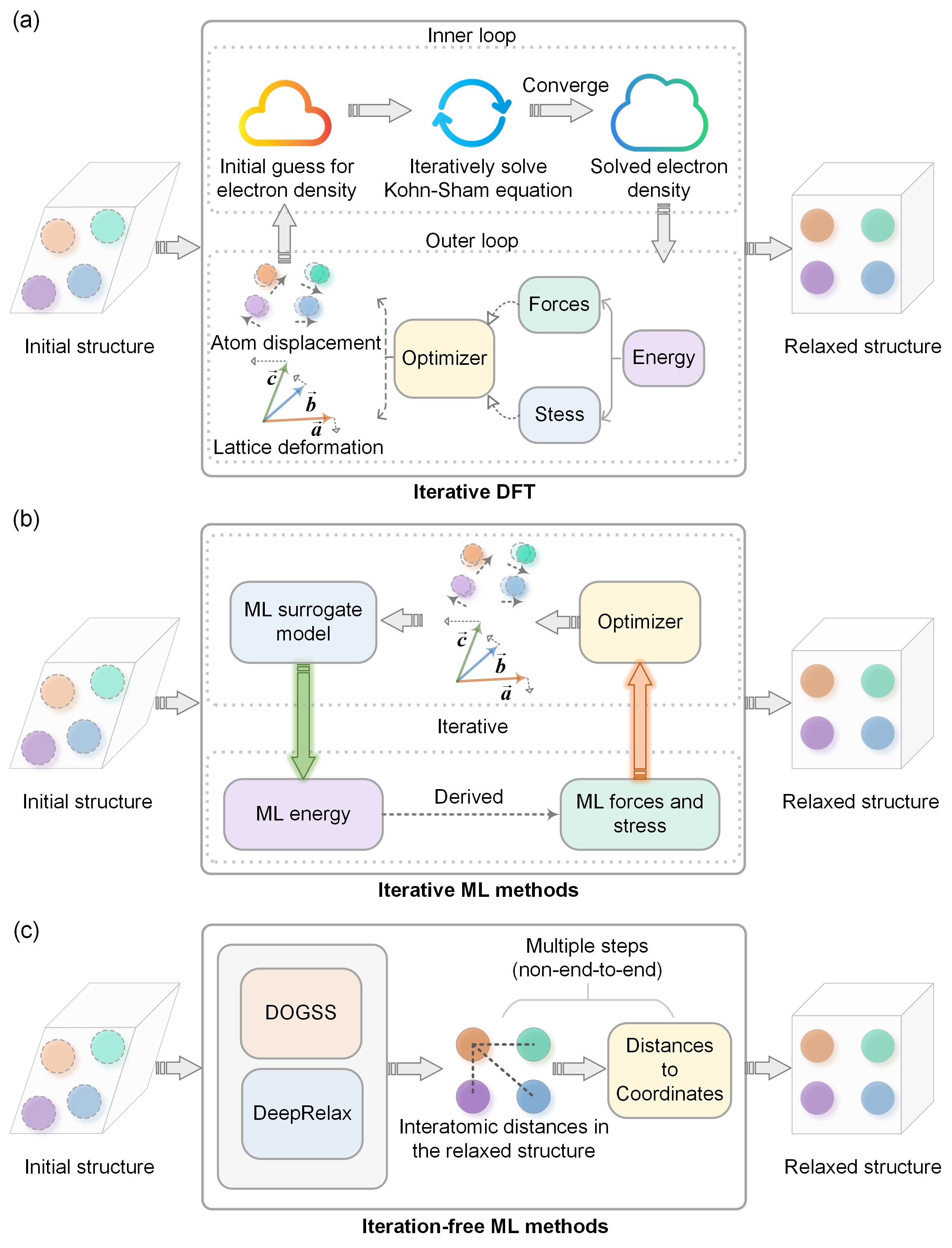}
  \caption{Structural optimization using DFT and ML methods. (a) DFT-based structural optimization involves both inner and outer loops. (b) Iterative ML models surrogate replaces the DFT inner loop but still runs a geometry‑update outer loop until forces converge. (c) Existing iteration-free ML models initially estimate interatomic distances, subsequently identifying the structure that aligns with these predicted distances via an optimization process. }
  \label{fgr:struct_relax}
\end{figure}

Structural optimization (or relaxation), which yields the lowest energy state or the most stable atomic configuration of a material (commonly called the relaxed structure), is a fundamental step in computational materials science because material properties are typically calculated based on relaxed structures. Traditionally, this process is achieved using \textit{ab initio} methods, particularly density functional theory (DFT) \cite{gibson2022data, lyngby2022data}. However, DFT calculations are computationally intensive due to the iterative procedures involved in two nested loops, as depicted in Figure \ref{fgr:struct_relax}(a). The inner loop solves the Kohn–Sham equations to self‐consistency, yielding the total energy and atomic forces for a fixed geometry. The outer loop then moves each atom according to those forces and repeats the inner loop until the forces fall below a convergence threshold. Together, these loops drive the structure to its minimum‐energy configuration.

To overcome these limitations, there has been growing interest in alternative methods, particularly machine learning (ML), for predicting relaxed structures more efficiently. ML methods for structural optimization can be broadly categorized into two types: iterative approaches \cite{chen2022universal, deng2023chgnet, batatia2022mace, gasteiger_dimenet_2020, mosquera2024machine, jiang2024machine, yang2025efficient} and iteration-free methods \cite{kim2023structure, yoon2020differentiable, yang2024scalable}. Iterative ML methods simplify the DFT calculation by eliminating the inner loop while retaining the outer loop calculations. At each step, a GNN‐based interatomic potential predicts the energy, forces, and stresses for the current atomic configuration; these quantities then drive a geometry update, after which the updated structure is fed back into the model for a new prediction, as illustrated in Figure \ref{fgr:struct_relax}(b). Despite their efficiency gains, iterative ML approaches have two key limitations. First, the development of iterative approaches heavily relies on the availability of comprehensive databases with detailed labels for energy, forces, and stresses during structural optimizations. Second, although these methods eliminate the inner loop, the outer loop remains, meaning the optimization process is still iterative and thus limits parallel scalability. 

To alleviate these issues, iteration-free methods have been developed to predict relaxed structures directly from the input unrelaxed ones using GNNs, entirely bypassing iterative loops. For example, DeepRelax \cite{yang2024scalable} and DOGSS \cite{yoon2020differentiable} employ a two-stage scheme: they first predict the set of interatomic distances for the relaxed structure, then reconstruct the atomic coordinates by solving a Euclidean distance geometry problem (Figure \ref{fgr:struct_relax}(c)). These iteration-free approaches nonetheless exhibit two key limitations. First, they model atomic interactions via message passing but do not explicitly account for lattice deformations. However, structure optimization typically involves both atomic displacements and lattice deformations, and variations in lattice parameters can substantially influence material properties. Consequently, the absence of explicit lattice modeling limits the accuracy of current GNNs in predicting relaxed structures. Second, although they avoid both inner and outer loop iterations, these methods often use a multi-step workflow instead of a fully end-to-end mapping, which can accumulate errors and reduce overall efficiency. 

To address these challenges, we introduce E$^{3}$Relax, an equivariant graph neural network designed for the end-to-end prediction of relaxed structures in a single step, as shown in Figure \ref{fgr:E$^{3}$Relax}(a). E$^{3}$Relax offers three advantages: 
\begin{itemize}
    \item \textbf{Unified atomic and lattice modeling.} E$^{3}$Relax promotes both atoms and lattice vectors to graph nodes, each equipped with an invariant scalar and an equivariant 3D vector feature. This dual‑node design enables the network to capture atomic displacements and lattice deformations simultaneously, overcoming the limitation of earlier models that represent only atoms.
    \item \textbf{True end‑to‑end optimization.} E$^{3}$Relax achieves end-to-end trainability by directly learning the mapping from unrelaxed to relaxed structures. This single-step framework avoids the potential cumulative error of multi-step approaches, thereby improving performance and efficiency.
    \item \textbf{Progressive structure refinement.} E$^{3}$Relax utilizes layer‐wise supervision to control the refinement process of atomic coordinates and lattice vectors throughout its network layers. This layer‐by‐layer guidance allows the model to progressively refine its predictions by mirroring the incremental convergence of DFT optimizations within a single, iteration‐free pass. We empirically demonstrate that this strategy yields substantial performance gains.
\end{itemize}

\begin{figure}[t]
  \centering
  \includegraphics[width=0.9\columnwidth]{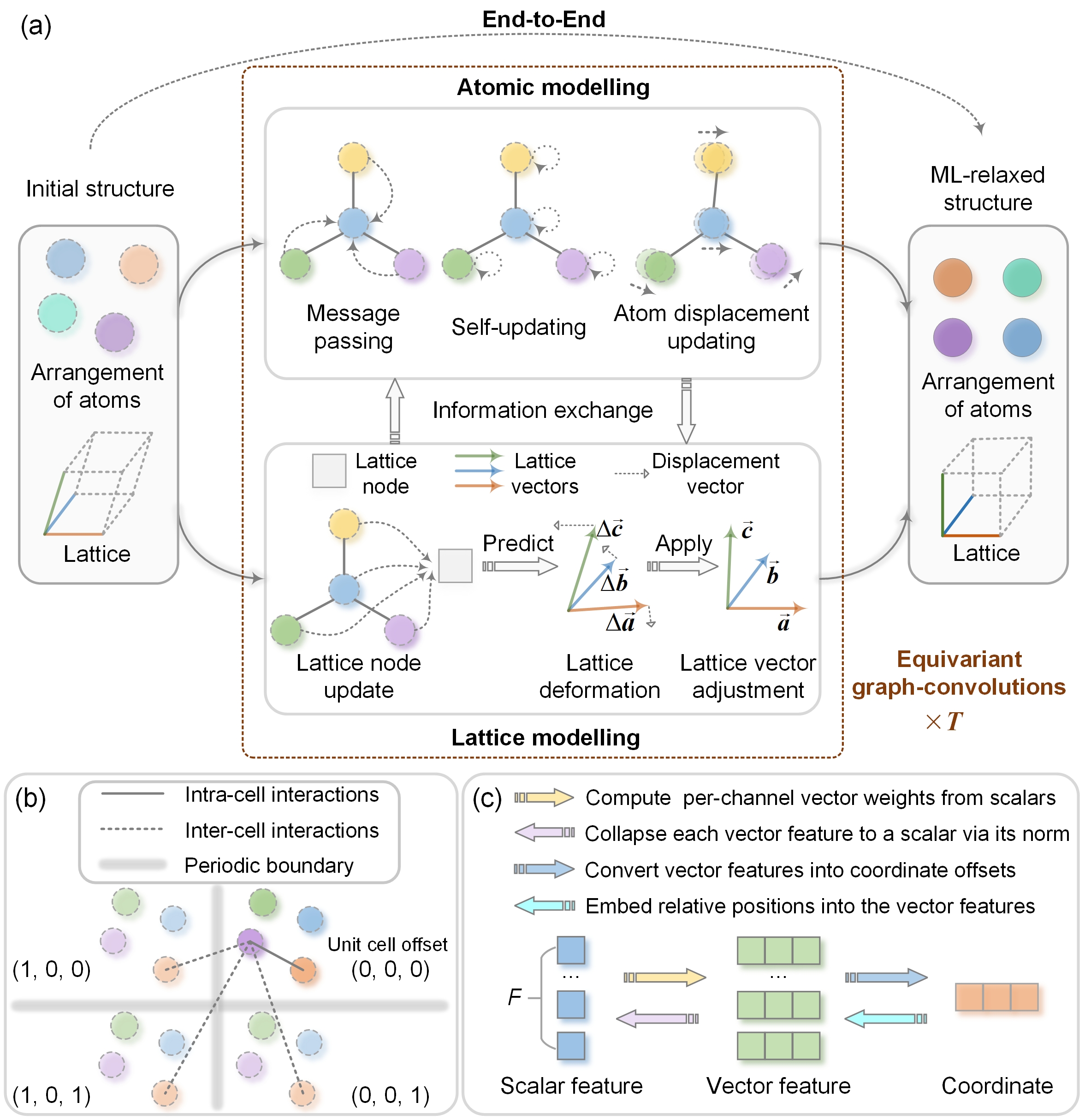}
  \caption{An overview of E$^{3}$Relax. (a) The model jointly captures atomic displacements and lattice deformations during structural optimization by modeling both atoms and the lattice vectors. (b) Illustration of a multi-edge graph showing an atom connected to the same neighboring atoms in different translated unit cells. (c) Depiction of the three varieties of node features along with their interrelations.}
  \label{fgr:E$^{3}$Relax}
\end{figure}

\section{Related Work}
\subsection{Iterative ML Models for Structural Optimization}
Iterative ML methods streamline DFT optimization by replacing the self-consistency inner loop with a GNN that iteratively predicts energies, forces, and stresses, while still retaining the outer geometry-update loop. The central challenge in these methods is designing GNN architectures that accurately capture atomic interactions. To ensure consistent energy predictions under translations, rotations, and reflections, many previously state-of-the-art (SOTA) models enforce invariance to the Euclidean group E(3) by relying on features such as bond lengths and angles \cite{xie2018crystal, schutt2017schnet, deng2023chgnet, gasteiger_dimenet_2020, choudhary2021atomistic, chen2022universal, gasteiger2021gemnet, Omee2024, li2025critical}. More recently, equivariant networks, which explicitly incorporate crystal symmetries and provide a richer geometric description, have demonstrated superior accuracy \cite{batatia2022mace, batzner20223, xu2024equivariant, wang2025elora, park2024scalable, schutt2021equivariant, deng2023chgnet}. However, iterative ML potentials still rely on extensive energy, force, and stress labels and retain an outer optimization loop, making the process inherently sequential and limiting parallel scalability.

\subsection{Iteration-Free Models for Structural Optimization}
Iteration-free methods bypass both the inner and outer optimization loops entirely \cite{yang2024scalable, yoon2020differentiable, kim2023structure, yang2025modeling}, aiming to directly map an unrelaxed structure to its relaxed counterpart without iterative computation. However, current SOTA models, such as DeepRelax \cite{yang2024scalable}, DOGSS \cite{yoon2020differentiable}, and Cryslator \cite{kim2023structure}, typically adopt multi-step pipelines, making them non-end-to-end. Crucially, all existing iteration‑free models focus exclusively on atoms, without providing an explicit representation for the lattice vectors. Since the cell’s geometry (its lattice parameters) directly affects properties like electronic band structure, density, and mechanical behavior, leaving out an explicit lattice representation may lead to less accurate or physically inconsistent predictions.

\section{Methodology}
Figure \ref{fgr:E$^{3}$Relax}(a) summarizes the E$^{3}$Relax workflow. Starting from an unrelaxed crystal defined by its atomic coordinates $\vec{\bm{r}}^{(0)}$ and lattice vectors $\vec{\bm{l}}^{(0)}$, the network applies a sequence of equivariant graph‑convolution layers. After the $t$-th layer the structure is updated to $\vec{\bm{r}}^{(t)}$ and $\vec{\bm{l}}^{(t)}$; successive layers therefore refine both atomic positions and lattice parameters until the final layer outputs $\vec{\bm{r}}^{(T)}$ and $\vec{\bm{l}}^{(T)}$. 

The key challenge is to maintain strict SE(3)-equivariance at every layer so that rotations or translations of the input produce identically transformed outputs for both atoms and lattice vectors. Preserving this symmetry is essential for physical correctness, robust generalization, and orientation‑independent predictions \cite{batatia2022mace, batzner20223, satorras2021n, schutt2021equivariant}. We achieve it by representing each atom and lattice vector with a paired scalar–vector feature set and by constraining all network operations to be SE(3)-equivariant.

Each equivariant graph‑convolution layer performs four steps: (1) message passing, during which a node gathers information from its neighbors to mimic atomic interactions; (2) self‑updating, which refines the node’s internal features; (3) atom position updating; and (4) lattice updating, which jointly adjusts atomic coordinates and lattice vectors toward lower energy. An additional lattice-atom interaction module enables the model to capture the coupling between atomic displacements and lattice deformations. All operations are SE(3)‑equivariant, ensuring that atomic coordinates and lattice vectors transform consistently under rotation and translation.

\subsection{Materials Graph Representation}
A crystal is encoded as a multi–edge graph  
\(\mathcal G=(\mathcal V^{a},\mathcal V^{l},\mathcal E)\),  
where \(v_i\!\in\!\mathcal V^{a}\;(i=1\ldots M)\) are atom nodes and  
\(v_c\!\in\!\mathcal V^{l}\;(c=1,2,3)\) are lattice nodes corresponding to the three lattice vectors. Unlike traditional materials graphs, which include only atom nodes, this dual-node formulation makes the lattice an explicit part of the graph structure.

Edges \(e_{ij}\!\in\!\mathcal E\) connect each atom to its \(K\) nearest neighbours within a cut‑off \(D\). Under periodic boundary conditions, the same atom pair may be connected by multiple edges, each annotated with a unit-cell offset \((k_1,k_2,k_3)\) corresponding to a translation \(k_1\vec{\bm{l}}_1 + k_2 \vec{\bm{l}}_2 + k_3 \vec{\bm{l}}_3\) (Figure~\ref{fgr:E$^{3}$Relax}(b)), where \(\vec{\bm l}_1,\vec{\bm l}_2,\vec{\bm l}_3\) are the three primitive lattice vectors. Every lattice node is linked to all atom nodes in the unit cell. Inspired by Matformer~\cite{yan2022periodic}, we further connect each atom node to its images in the six neighbouring cells—offsets \((0,0,1)\), \((0,1,0)\), \((1,0,0)\), \((0,1,1)\), \((1,0,1)\), and \((1,1,0)\)—to expose long-range periodicity, which we find consistently improves performance.

Each atom node carries an invariant scalar \(\bm x_i\in \mathbb{R}^{F}\), an equivariant vector \(\vec{\bm x}_i\in\mathbb {R}^{F\times3}\), and its position \(\vec{\bm r}_i\in\mathbb {R}^{3}\). Each lattice node is also assigned three distinct features, denoted as $\bm{y}_c \in \mathbb{R}^F$, $\vec{\bm{y}}_c \in \mathbb{R}^{F \times 3}$, and $\vec{\bm{l}}_c \in \mathbb{R}^3$. Initialisation details of these features are provided in Appendix A. The interactions among these feature types are illustrated in Figure \ref{fgr:E$^{3}$Relax}(c). For each atom pair we also define the relative vector \(\vec{\bm{r}}_{ij} = \vec{\bm{r}}_j - \vec{\bm{r}}_i\)  to encode local geometry.

\subsection{Message Passing}
Message passing is a method by which an atom node $v_i$ collects information from its neighboring nodes $v_j$. In this work, we implement the message passing as proposed by PAINN \cite{schutt2021equivariant}. In the $t$-th layer, during message passing, each node $v_i$ aggregates messages from the scalar features $\bm{x}_j^{(t)}$ and vector features $\vec{\bm{x}}_j^{(t)}$ of its neighboring nodes $v_j$. This aggregation forms intermediate scalar and vector variables $\bm{m}_i$ and $\vec{\bm{m}}_i$, respectively, calculated as follows:
\begin{equation}
    \bm{m}_i = \sum_{v_j \in \mathcal{N}(v_i)} \phi_h (\bm{x}_{j}^{(t)}) \circ \gamma_h \big( \lambda (\Vert \vec{\bm{r}}_{ji}^{(t)} \Vert) \big)
\end{equation}
\begin{align}
    \vec{\bm{m}}_i = \sum_{v_j \in \mathcal{N}(v_i)} & \phi_u (\bm{x}_{j}^{(t)}) \circ \gamma_u \big(\lambda(\Vert \vec{\bm{r}}_{ji}^{(t)} \Vert) \big) \circ \vec{\bm{x}}_j^{(t)} \nonumber \\
    & + \phi_v (\bm{x}_{j}^{(t)}) \circ \gamma_v \big(\lambda(\Vert \vec{\bm{r}}_{ji}^{(t)} \Vert)\big) \circ \frac{\vec{\bm{r}}_{ji}^{(t)}}{\Vert \vec{\bm{r}}_{ji}^{(t)} \Vert}
\end{align}
Here, $\mathcal{N}(v_i)$ denotes the set of neighboring nodes to $v_i$. The functions $\phi_h, \phi_u, \phi_v: \mathbb{R}^{F} \rightarrow \mathbb{R}^{F}$ are multilayer perceptrons (MLPs), $\lambda$ represents $K$ Gaussian radial basis functions (RBF) \cite{schutt2017schnet} used to expand bond distances, and $\gamma_h, \gamma_u, \gamma_v: \mathbb{R}^{K} \rightarrow \mathbb{R}^{F}$ are also MLPs.

\subsection{Self-Updating}
Self‑updating leverages each node’s internal context to fuse scalar and vector features. This process combines the $F$ scalar and $F$ vector elements within $\bm{m}_i$ and $\vec{\bm{m}}_i$ to generate updated scalar $\bm{x}_i^{(t+1)}$ and vector $\vec{\bm{x}}_i^{(t+1)}$ as follows:
\begin{equation}
    \bm{x}_i^{(t+1)} = \phi_{s1}(\bm{m}_{i} \oplus \Vert \bm{U} \vec{\bm{m}}_i \Vert) + \tanh\big(\phi_{s2}(\bm{m}_{i} \oplus \Vert \bm{U} \vec{\bm{m}}_i \Vert)\big)\bm{m}_i
\end{equation}
\begin{equation}
    \vec{\bm{x}}_i^{(t+1)} = \phi_{v}(\bm{m}_{i} \oplus \Vert \bm{U} \vec{\bm{m}}_i \Vert) \circ (\bm{V} \vec{\bm{m}}_i)
\end{equation}
where $\oplus$ denotes concatenation, $\phi_{s1}, \phi_{s2}, \phi_{v}: \mathbb{R}^{2F} \rightarrow \mathbb{R}^F$ are MLPs, and $\bm{U}, \bm{V} \in \mathbb{R}^{F \times F}$ are trainable matrices. The $\tanh\big( \cdot \big)$ function serves as a gate, regulating how much information from $\bm{m}_i$ is preserved.

\subsection{Atom Position Updating}
Atom position updating refines the atomic positions to mirror the structure optimization process. Initially, we compute the coefficients $\alpha_{i}^f$ for each vector $\vec{\bm{x}}_i^f \in \vec{\bm{x}}_i$ (i.e., $\vec{\bm{x}}_i^f \in \mathbb{R}^{1 \times 3}$ represents one of the $F$ vectors in $\vec{\bm{x}}_i$):
\begin{equation}
    \bm{p}_i = g(\bm{x}_i)
\end{equation}
\begin{equation}
    \alpha_{i}^f = \frac{\exp{(p_i^f)}}{\sum_k \exp{(p_i^k)}}
\end{equation}
where $g: \mathbb{R}^F \rightarrow \mathbb{R}^F$ is an MLP. The positions of the atoms are updated as follows:
\begin{equation}
    \Delta \vec{\bm{r}}_i = \bm{W}\vec{\bm{x}}_i + \sum_f \alpha_{i}^f\vec{\bm{x}}_i^f
\end{equation}
\begin{equation}
    \vec{\bm{r}}^{(t+1)}_i = \vec{\bm{r}}^{(t)}_i + \Delta \vec{\bm{r}}_i
\end{equation}
where $\bm{W} \in \mathbb{R}^{1 \times F}$ is a trainable matrix.

\subsection{Lattice Updating}
In most crystal GNNs the lattice vectors appear only implicitly: they determine the periodic image offsets used to build the multi‑edge graph, but they are not represented as learnable entities, nor are they updated during message passing. To overcome this limitation, we use lattice nodes to represent the lattice vectors. These nodes carry their own features and are refined layer‑by‑layer alongside the atom nodes.

Specifically, lattice updating begins by first updating the lattice node representations, followed by predicting the displacement vectors and applying these to adjust the lattice vectors. Particularly, we update the lattice node representations $\bm{y}_c^{(t)}$ and $\vec{\bm{y}}_c^{(t)}$ to intermediate states $\bm{n}_c$ and $\vec{\bm{n}}_c$ using the following equations:
\begin{equation}
    \bm{n}_c = \gamma \left( \left(\frac{1}{M} \sum_{v_j \in \mathcal{V}^a} \bm{x}_{j}^{(t)} \right) \oplus \bm{y}_c^{(t)} \right)
\end{equation}
\begin{equation}
    \vec{\bm{n}}_c = \bm{W} \left( \left(\frac{1}{M} \sum_{v_j \in \mathcal{V}^a} \vec{\bm{x}}_{j}^{(t)} \right) + \vec{\bm{y}}_c^{(t)} \right)
\end{equation}
where $\gamma: \mathbb{R}^{2F} \rightarrow \mathbb{R}^F$ is an MLP, and $\bm{W} \in \mathbb{R}^{F\times F}$ is a trainable matrix. 

Next, the updated lattice node representations $\bm{y}_c^{(t+1)}$ and $\vec{\bm{y}}_c^{(t+1)}$ are computed as:
\begin{equation}
    \bm{y}_c^{(t+1)} = \phi_{s1}(\bm{n}_c \oplus \Vert \bm{U} \vec{\bm{n}}_c \Vert) + \tanh\left(\phi_{s2}(\bm{n}_c \oplus \Vert \bm{U} \vec{\bm{n}}_c \Vert)\right)\bm{n}_c
\end{equation}
\begin{equation}
    \vec{\bm{y}}_c^{(t+1)} = \phi_{v}(\bm{n}_c \oplus \Vert \bm{U} \vec{\bm{n}}_c \Vert) \circ (\bm{V} \vec{\bm{n}}_c) + \vec{\bm{n}}_c
\end{equation}
where $\phi_{s1}, \phi_{s2}, \phi_{v}: \mathbb{R}^{2F} \rightarrow \mathbb{R}^{F}$ are MLPs, and $\bm{U}, \bm{V} \in \mathbb{R}^{F \times F}$ are trainable matrices. 

Finally, the lattice coordinates are updated as:
\begin{equation}
    \Delta \vec{l}_c = \bm{W}_p \vec{\bm{y}}_c^{(t+1)} 
\end{equation}
\begin{equation}
    \vec{\bm{l}}_c^{(t+1)} = \vec{\bm{l}}_c^{(t)} + \Delta \vec{l}_c
\end{equation}
where $\bm{W}_p \in \mathbb{R}^{1 \times F}$ is a trainable matrix, and $\Delta \vec{l}_c$ represents the predicted displacement vector (lattice deformation).

\subsection{Lattice–Atom Interaction}
The lattice updating step described above transfers information only from atom nodes to lattice nodes; it does not allow information to flow back from the lattice to the atoms. To remedy this, we introduce a lattice–atom interaction module that broadcasts lattice information to every atom. For each lattice node \(v_c\in\mathcal V^l\), we first compute intermediate atom node features \(\bm n_i\) and \(\vec{\bm n}_i\) by incorporating the relative distance between atom node \(v_i \in\mathcal V^a\) and lattice node \(v_c \in\mathcal V^l\):
\begin{equation}
    \bm{n}_i =  \phi_h (\bm{x}_i^{(t)}) \circ \gamma_h \left(\lambda \left( \Vert \vec{\bm{r}}_i^{(t)} - \vec{\bm{l}}_c^{(t)} \Vert \right) \right) + \bm{x}_i^{(t)}
\end{equation}
\begin{align}
    \vec{\bm{n}}_i = & \phi_u (\bm{x}_i^{(t)}) \circ \gamma_u \left( \lambda \left( \Vert \vec{\bm{r}}_i^{(t)} - \vec{\bm{l}}_c^{(t)} \Vert \right) \right) \circ \vec{\bm{x}}_i^{(t)} \nonumber \\
    & + \phi_v (\bm{x}_i^{(t)}) \circ \gamma_v \left(  \lambda \left( \Vert \vec{\bm{r}}_i^{(t)} - \vec{\bm{l}}_c^{(t)} \Vert \right) \right) \circ \frac{\left(\vec{\bm{r}}_i^{(t)} - \vec{\bm{l}}_c^{(t)}\right)}{\Vert \vec{\bm{r}}_i^{(t)} - \vec{\bm{l}}_c^{(t)} \Vert}
\end{align}
Here, $\phi_h, \phi_u, \phi_v: \mathbb{R}^{F} \rightarrow \mathbb{R}^{F}$ are MLPs, $\lambda$ represents a set of RBF used to expand bond distances, and $\gamma_h$, $\gamma_u$, and $\gamma_v: \mathbb{R}^{2F} \rightarrow \mathbb{R}^{F}$ are also MLPs. We apply this update once per lattice node \(c=1,2,3\). Finally, we fuse the intermediate features \(\bm n_i\) and \(\vec{\bm n}_i\) with the current lattice node feature \(\bm y_c^{(t)},\vec{\bm y}_c^{(t)}\) to obtain the next-layer atom representations:
\begin{equation}
    \bm{x}_i^{(t+1)} = f \left(\bm{n}_i \oplus \bm{y}_c^{(t)}\right) + \bm{n}_i
\end{equation}  
\begin{equation}
    \vec{\bm{x}}_i^{(t+1)} = \bm{W}_v \left(\vec{\bm{n}}_i + \vec{\bm{y}}_c^{(t)}\right) + \vec{\bm{n}}_i
\end{equation}
where $f: \mathbb{R}^{2F} \rightarrow \mathbb{R}^F$ is an MLP and $\bm{W}_v \in \mathbb{R}^{F \times F}$ is a trainable matrix. Again, we apply this update once per lattice node \(c=1,2,3\).

\subsection{Layer-Wise Supervision}
If supervision is applied only to the final output, the network may simply copy its intermediate representations forward and perform one large correction in the last layer. Such abrupt “jumps’’ are hard to learn. Instead, E$^{3}$Relax attaches a loss to every layer so that each depth must contribute a small, physically meaningful correction. In other words, the model is trained to progressively approach the same optimal configuration across all depths. The overall loss is defined as:
\begin{equation}
    \mathcal{L} \;=\; \sum_{t=1}^{T} \alpha_t \Biggl(
      \sum_{v_i \in \mathcal{V}^a}\bigl\lvert \vec{\bm{r}}^{(t)}_i - \tilde{\bm{r}}_{i} \bigr\rvert
      \;+\;
      \sum_{v_c \in \mathcal{V}^l}\bigl\lvert \vec{\bm{l}}_c^{(t)} - \tilde{\bm{l}}_c \bigr\rvert
    \Biggr)
    \label{eq:loss}
\end{equation}
where $\alpha_t$ controls each layer’s contribution, and all $\alpha_t$ are set to 1 by default. 

\subsection{Equivariance Analysis}
In E$^{3}$Relax, all non-linear activations (MLPs, gating functions, etc.) are applied exclusively to scalar feature channels (e.g., $\bm{x}_{j}$), while every operation on vector features (e.g., $\vec{\bm{x}}_{j}$) is either a scaling (i.e., $\circ$) or linear combinations (e.g., $\bm{W}\vec{\bm{x}}_j$). By construction, following the equivariant message-passing framework of PAINN \cite{schutt2021equivariant}, these components guarantee exact SE(3)-equivariance throughout the network.

\subsection{Implementation Details}
The E$^{3}$Relax model is implemented using PyTorch and experiments are conducted on an NVIDIA L40S GPU with 48 GB of memory. We employ the AdamW optimizer with an initial learning rate of 0.0001 to update the model parameters. Besides, we incorporate a learning rate decay strategy, which reduces the learning rate if no improvement is observed in a specific performance metric over a period of 5 epochs.

\section{Experiments}
\subsection{Datasets}
We mainly use four datasets in this study: Materials Project (MP) \cite{jain2013commentary}, X-Mn-O (X = Mg, Ca, Ba, Sr) \cite{kim2023structure}, C2DB \cite{gjerding2021recent, lyngby2022data}, and Layered van der Waals (vdW) crystals \cite{yang2024scalable}. Each data point comprises a pair of structures, i.e., one unrelaxed and one DFT-relaxed. These datasets include both 2D and 3D materials and represent a wide range of compositions and structural complexities, making them well-suited benchmarks for evaluating structure optimization models. Detailed descriptions of these datasets are provided in Appendix B.

\subsection{Baselines}
Consistent with the protocol used in DeepRelax \cite{yang2024scalable}, we compare E$^{3}$Relax with SOTA iteration-free equivariant models, including DeepRelax \cite{yang2024scalable}, EquiformerV2 \cite{equiformer_v2}, HEGNN \cite{cen2024high}, GotenNet \cite{aykent2025gotennet}, PAINN \cite{schutt2021equivariant}, and EGNN \cite{satorras2021n}. Each baseline is evaluated using the authors’ original implementations and default hyperparameter settings. All models are evaluated on identical training, validation, and testing sets. The Dummy model, which simply returns the input initial structure as the output, served as a control in our assessments. Detailed information on baseline implementations can be found in Appendix C.

\subsection{Performance Indicators}
\label{sec:metrics}
We use three metrics to quantify the discrepancy between predicted and DFT-relaxed structures: mean absolute error (MAE) of atomic coordinates, cell shape deviation, and MAE of cell volume. The coordinate MAE evaluates positional errors, while cell shape deviation and volume MAE assess lattice geometry errors. Detailed definitions and computation procedures are provided in Appendix~D.  

\begin{table*}[!ht]
\centering
\begin{tabular}{lcccccc}
\toprule
\multirow{2}{*}{Model} &
\multicolumn{3}{c}{MP} &
\multicolumn{3}{c}{X-Mn-O} \\ 
\cmidrule(lr){2-4}\cmidrule(lr){5-7}
& MAE Coords.$\downarrow$ & Shape Dev.$\downarrow$ & MAE Vol.$\downarrow$
& MAE Coords.$\downarrow$ & Shape Dev.$\downarrow$ & MAE Vol.$\downarrow$ \\ \midrule
Dummy              & 0.095 & 11.502 & 27.006 & 0.314 & 13.913 & 32.839 \\
PAINN              & 0.088 &  4.775 &  9.343 & 0.159 &  3.784 &  3.803 \\
EGNN               & 0.086 &  4.834 &  9.253 & 0.166 &  3.814 &  4.208 \\
HEGNN              & 0.091 &  4.987 & 10.382 & 0.191 &  3.862 &  4.421 \\
GotenNet           & 0.078 &  \underline{4.111} &  \underline{7.456} & 0.160 &  3.669 &  3.559 \\
EquiformerV2       & 0.069 &  4.329 &  8.539 & \underline{0.115} &  3.558 &  3.713 \\
DeepRelax          & \underline{0.066} &  4.554 &  9.611 & 0.116 &  \underline{3.530} &  \underline{3.442} \\
E$^{3}$Relax       & \textbf{0.057} & \textbf{4.020} & \textbf{7.417}
                   & \textbf{0.105} & \textbf{3.447} & \textbf{3.369} \\ 
Improvement        & 13.64\% & 2.21\% & 0.52\% & 8.70\% & 2.35\% & 2.12\% \\
\bottomrule
\end{tabular}
\caption{Comparison of E$^{3}$Relax with baseline models on the two 3D materials datasets (MP and X–Mn–O). Metrics are MAE of atomic coordinates ($\mathrm{\AA}$), cell shape deviation ($\mathrm{\AA}$), and MAE of cell volume ($\mathrm{\AA^{3}}$) between ML-relaxed and DFT-relaxed structures. The “Improvement” rows report the percentage gain of E$^{3}$Relax over the second-best model for each metric. The best results are in bold, and the second-best are underlined.}
\label{tbl:3D_performance}
\end{table*}

\subsection{Results on 3D Materials Datasets}
\begin{figure}[tb]
  \centering
  \includegraphics[width=0.9\columnwidth]{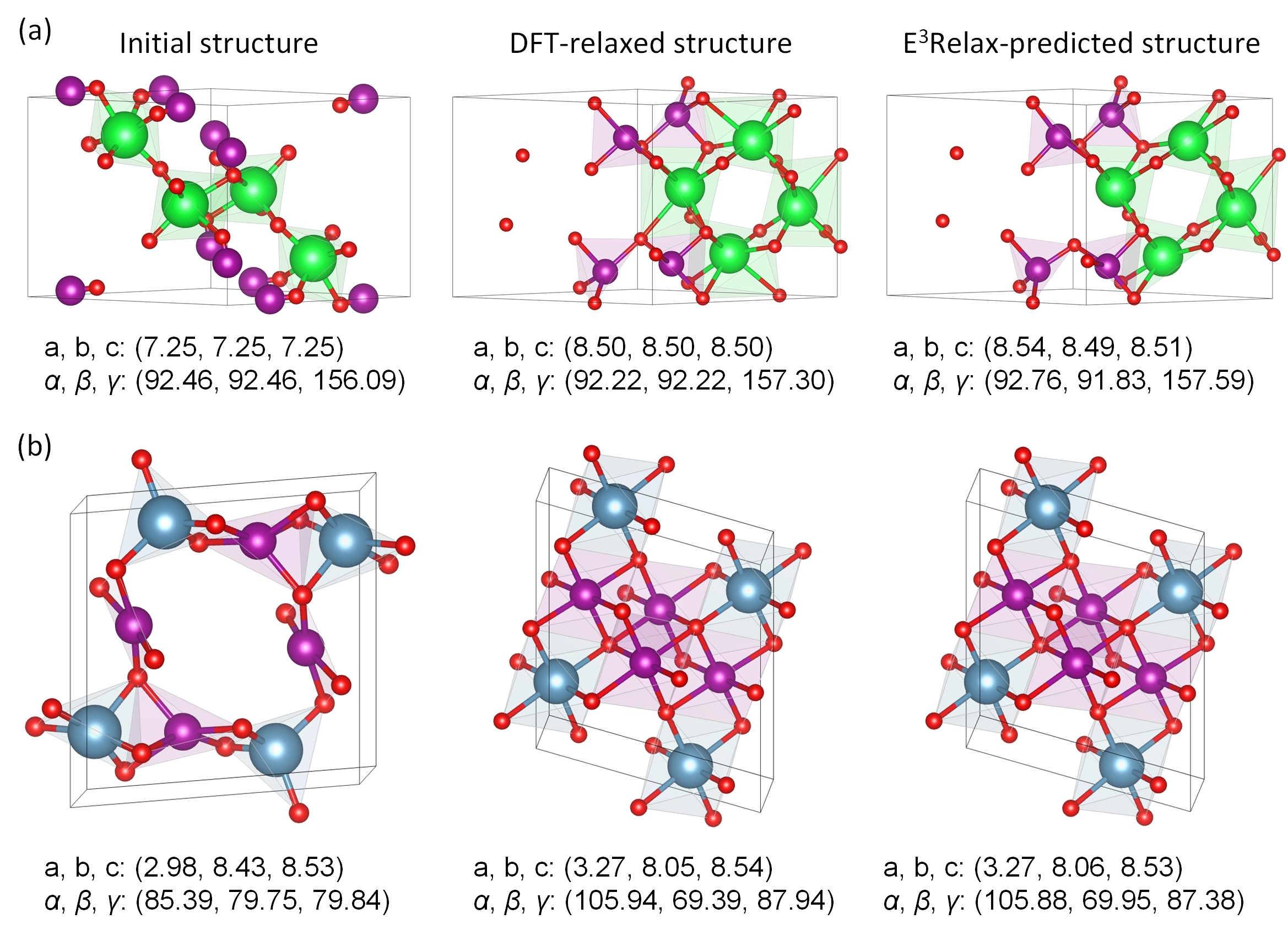}
  \caption{Two crystal structures from the X-Mn-O dataset optimized with E$^{3}$Relax: (a) $\rm Mn_4O_8Sr_4$ and (b) $\rm Ca_4Mn_4O_8$. \(a\), \(b\), and \(c\) are lattice constants in angstroms ($\mathrm{\AA}$), and \(\alpha\), \(\beta\), and \(\gamma\) are angles in degrees ($^\circ$).}
  \label{fgr:xmno}
\end{figure}

We first evaluate E$^{3}$Relax on the two 3D materials datasets. As shown in Table~\ref{tbl:3D_performance}, E$^{3}$Relax consistently surpasses previous SOTA models across all evaluation metrics. For a qualitative assessment of our model, we select two structures from X–Mn–O dataset, $\mathrm{Mn_4O_8Sr_4}$ and $\mathrm{Ca_4Mn_4O_8}$, and optimize them with E$^{3}$Relax (Figure~\ref{fgr:xmno}). As can be seen, the E$^3$Relax-predicted structures are highly consistent with the DFT-relaxed ones. Further analysis and visualizations are provided in Appendices E and F.

\begin{table*}[!ht]
\centering
\begin{tabular}{lcccccc}
\toprule
\multirow{2}{*}{Model} &
\multicolumn{3}{c}{C2DB} &
\multicolumn{3}{c}{Layered vdW} \\ 
\cmidrule(lr){2-4}\cmidrule(lr){5-7}
& MAE Coords.$\downarrow$ & Shape Dev.$\downarrow$ & MAE Vol.$\downarrow$
& MAE Coords.$\downarrow$ & Shape Dev.$\downarrow$ & MAE Vol.$\downarrow$ \\ \midrule
Dummy              & 0.268 & 10.615 & 149.648 & 0.103 & 0.172 & 2.991 \\
PAINN              & 0.226 &  5.952 &  61.897 & 0.035 & 0.294 & 2.782 \\
EGNN               & 0.232 &  6.065 &  67.892 & 0.033 & \underline{0.098} & 1.636 \\
HEGNN              & 0.244 &  6.329 & 71.625  & 0.049 & 0.312 & 4.989 \\
GotenNet           & 0.208 &  5.704 & \textbf{55.800}  & \underline{0.028} & 0.204 & \underline{1.006} \\
EquiformerV2       & \underline{0.188} &  \underline{5.660} &  60.435 & 0.034 & 0.228 & 3.617 \\
DeepRelax          & 0.196 &  5.752 &  60.216 & 0.052 & 0.340 & 1.166 \\
E$^{3}$Relax       & \textbf{0.171} & \textbf{5.430} & \underline{56.909}
                   & \textbf{0.025} & \textbf{0.087} & \textbf{0.871} \\ 
Improvement        & 9.04\% & 4.07\% & -1.99\% & 10.71\% & 11.22\% & 13.42\% \\
\bottomrule
\end{tabular}
\caption{Comparison of E$^{3}$Relax with baseline models on the two 2D materials datasets (C2DB and Layered vdW). }
\label{tbl:2D_performance}
\end{table*}

\subsection{Results on 2D Materials Datasets}
We further evaluate the model on 2D materials using the C2DB and layered vdW datasets. As reported in Table~\ref{tbl:2D_performance}, E$^{3}$Relax maintains strong performance and continues to outperform or remain competitive with existing approaches. Additional results and visualizations are available in Appendices E and F.

\subsection{Ablation Study}
The effectiveness of E$^{3}$Relax is attributed to three key strategies: First, its dual-node representation explicitly captures both atomic displacements and lattice deformations. Second, the prediction process is designed as a series of refinements of atomic coordinates and lattice vectors. Third, incorporating self-connecting edges into the graph representation enhances the model's ability to capture long-range periodicity. To evaluate the necessity of these approaches, we compared E$^{3}$Relax against three variants:
\begin{itemize}
    \item \textbf{w/o lattice nodes}: Here, lattice nodes are removed, and invariant scalar features $\bm{x}_i$ are employed for predicting relaxed lattice vectors, following the procedure outlined in DeepRelax \cite{yang2024scalable}.
    \item \textbf{w/o refinement}: This variant omits the progressive refinement process, updating atomic coordinates and lattice vectors only in the final layer without layer-wise supervision.
    \item \textbf{w/o self-connecting edges}: Omits the use of self-connecting edges.
\end{itemize}

Table \ref{tbl:ablation_3D} presents the experimental results on the MP and X-Mn-O datasets, showing that E$^{3}$Relax significantly outperforms the three variants. Notably, introducing lattice nodes with equivariant lattice–atom interactions yields substantial performance gains, underscoring the critical role of equivariance in accurately modeling coupled atomic–lattice dynamics.

We also conduct a sensitivity analysis of key hyperparameters, including the number of layers ($T$), hidden dimension size ($F$), and layer weights ($\alpha_t$), on the X–Mn–O dataset. The results confirm that E$^{3}$Relax is robust to variations in these hyperparameters, with performance remaining stable across different configurations (see Appendix~G).

\begin{table*}[!ht]
\centering
\begin{tabular}{lcccccc}
\toprule
\multirow{2}{*}{Model} &
\multicolumn{3}{c}{MP} &
\multicolumn{3}{c}{X–Mn–O} \\
\cmidrule(lr){2-4}\cmidrule(lr){5-7}
& MAE Coords.$\downarrow$ & Shape Dev.$\downarrow$ & MAE Vol.$\downarrow$
& MAE Coords.$\downarrow$ & Shape Dev.$\downarrow$ & MAE Vol.$\downarrow$ \\ \midrule
Dummy                         & 0.095 & 11.502 & 27.006 & 0.314 & 13.913 & 32.839 \\
w/o lattice nodes   & 0.075 &  4.389 &  7.964 & 0.158 &  3.576 &  3.634 \\
w/o refinement            & 0.075 &  4.565 &  8.465 & 0.146 &  3.581 &  3.747 \\
w/o self-connecting edges & 0.058 &  4.114 &  7.553 & 0.144 &  3.839 &  3.494 \\
E$^{3}$Relax                  & \textbf{0.057} & \textbf{4.020} & \textbf{7.417}
                              & \textbf{0.105} & \textbf{3.447} & \textbf{3.369} \\
\bottomrule
\end{tabular}
\caption{Ablation study on the two 3D materials datasets (MP and X–Mn–O). }
\label{tbl:ablation_3D}
\end{table*}

\subsection{Comparison with Iterative ML Models}
To benchmark E$^3$Relax against iterative ML-based potentials, we use a $\mathrm{MoS_2}$ dataset with point defects containing energy, force, and stress labels. This dataset, compiled by Huang et al. \cite{huang2023unveiling}, comprises 5933 unique low-defect configurations in an $8\times8$ supercell. We split the data into 4746 training, 593 validation, and 594 test samples and train two state-of-the-art interatomic potentials, M3GNet \cite{chen2022universal} and CHGNet \cite{deng2023chgnet} (details in Appendix H), using the same splits.

\begin{figure}[!t]
  \centering
  \includegraphics[width=0.9\columnwidth]{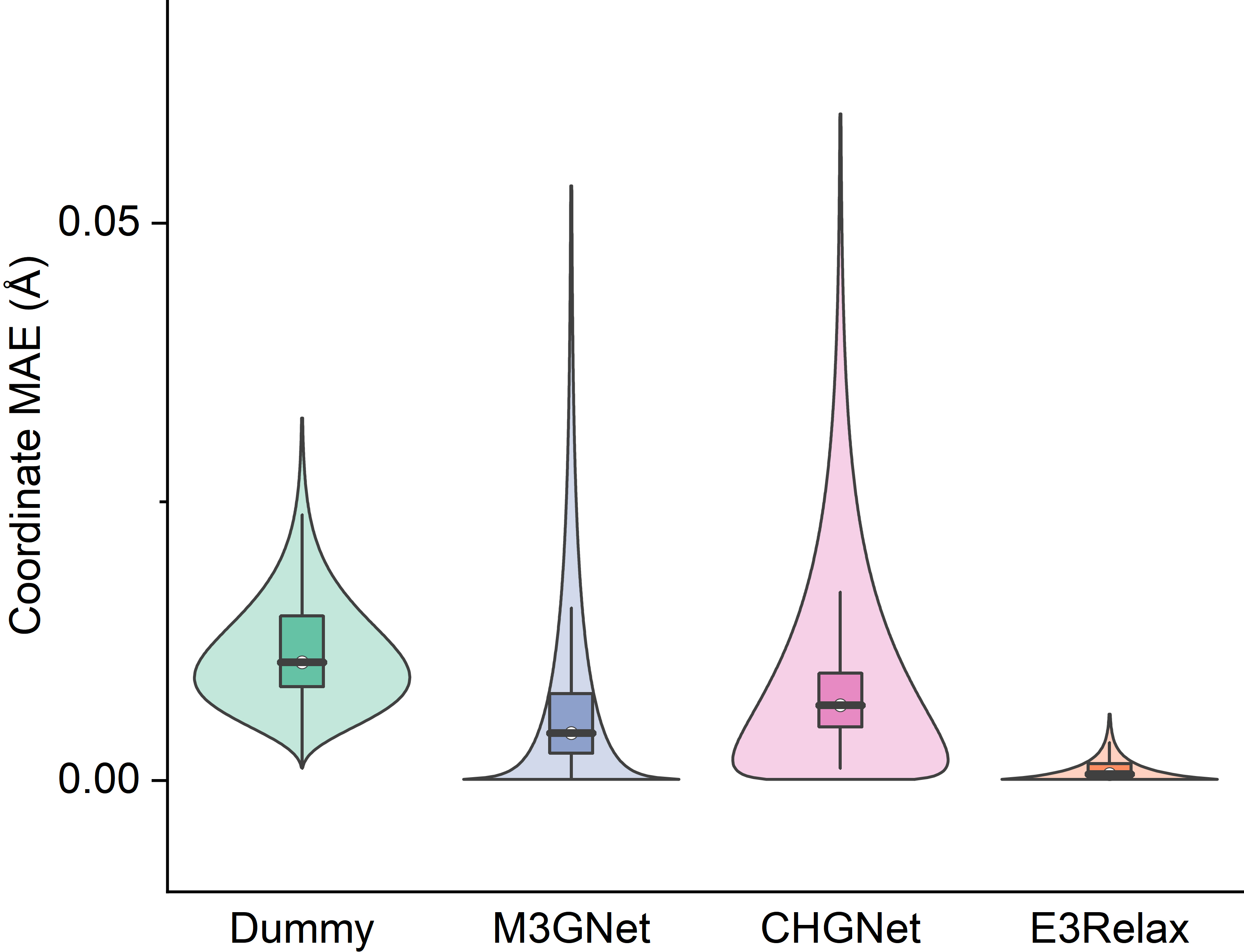}
  \caption{Comparison of coordinate MAE ($\rm \AA$) between E$^3$Relax and two ML-potential models using violin plots.}
  \label{fgr:compare_ml_pot}
\end{figure}

Because all structures share identical lattice parameters before and after optimization, we report only the coordinate MAE. As shown in Figure \ref{fgr:compare_ml_pot}, E$^3$Relax notably outperforms both M3GNet and CHGNet. We also observe that the iterative methods exhibit larger standard deviations in their coordinate errors, likely due to error accumulation over iterative optimization steps.

\subsection{DFT Validation}
\label{sec:dft_val}
\begin{figure}[ht]
  \centering
  \includegraphics[width=0.9\columnwidth]{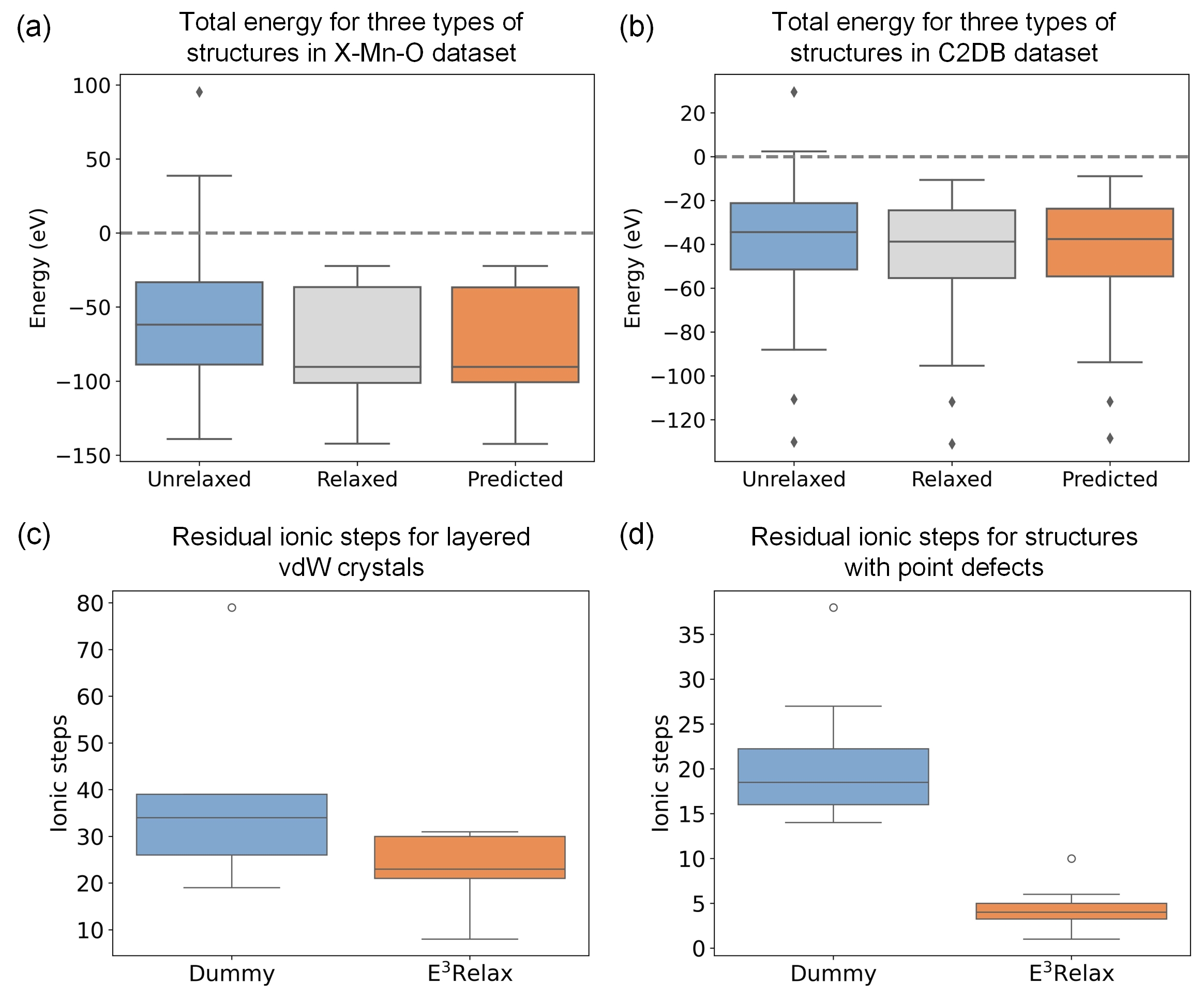}
  \caption{Box plots of total energy distributions and residual ionic steps for E$^{3}$Relax predictions.
(a) and (b) compare total energies of unrelaxed, DFT‐relaxed, and E$^{3}$Relax‐predicted structures on the X–Mn–O and C2DB datasets, respectively.
(c) and (d) report the number of residual ionic steps required to converge DFT optimizations when initialized from unrelaxed and E$^{3}$Relax‐predicted structures for the layered vdW crystal dataset and the MoS\textsubscript{2} defect dataset, respectively }
  \label{fgr:DFT_AAAI}
\end{figure}

To further validate E$^{3}$Relax, we compute total energies for three sets of structures, unrelaxed, DFT-relaxed, and E$^{3}$Relax‐predicted, to determine whether our predictions are energetically competitive. Detailed DFT parameters are provided in Appendix I. For this analysis, we randomly sampled 100 test‐set structures each from the X–Mn–O (3D materials) and C2DB (2D materials) datasets. Figures \ref{fgr:DFT_AAAI}(a) and (b) compare the energy distributions among the three types of structures: the median and spread of the E$^{3}$Relax‐predicted structures closely match those of the fully DFT‐relaxed structures, demonstrating our model’s ability to identify near-ground-state configurations.

We further assess how well E$^{3}$Relax predictions accelerate DFT optimizations by measuring the residual number of ionic steps required when using our predicted structures as starting points. We performed this test on the layered vdW crystal dataset and on the MoS$_2$ point‐defect dataset (for which we randomly sampled 20 test structures to limit computational cost). Figures \ref{fgr:DFT_AAAI}(c) and (d) show that, on average, initializing DFT optimization from E$^{3}$Relax outputs substantially reduces the number of ionic steps needed to converge.

\subsection{Complexity Analysis}

For a crystal containing \(N\) atoms with \(K\) neighbours each and a hidden feature size \(F\), lightweight equivariant GNNs such as PAINN or EGNN incur a per-layer cost of \(\mathcal O(N K F^2)\).  The extra cost of E$^{3}$Relax mainly comes from two additional modules: the lattice updating block, costing \(\mathcal O(3N F^2 + F^2)\), and the lattice–atom interaction block, costing \(\mathcal O(3N F^2)\).  Combining these with the base message‐passing and self‐update operations yields a total per‑layer complexity of $\mathcal O\bigl(N (K+6) F^2)$. When \(K\approx 50\) (typical setting), the extra \(6N F^2\) overhead represents only about a 12\% increase in compute.

We benchmarked inference times on an NVIDIA L40S GPU for two iterative relaxers (CHGNet and M3GNet), a two-stage single‐shot model (DeepRelax), and E$^{3}$Relax.  Averaged over structures in the point defect dataset, CHGNet required \(3.72\pm5.13\) s, M3GNet \(6.64\pm8.01\) s, DeepRelax \(0.47\pm0.04\) s, and E$^{3}$Relax only \(0.021\pm0.001\) s per structure. This speed‑up stems from E$^{3}$Relax’s iteration‑free, single‑step architecture, which eliminates the repeated optimization or multi‑stage calls inherent in the other methods.

\subsection{Conclusion}
We have presented E$^{3}$Relax, an iteration-free, equivariant graph neural network that directly predicts relaxed crystal structures in a single end-to-end step. By promoting both atoms and lattice vectors to learnable graph nodes with dual scalar-vector features, E$^{3}$Relax captures local displacements and global lattice deformations in a unified, symmetry‑preserving framework. Across four benchmark datasets, including both 2D and 3D materials, E$^{3}$Relax consistently outperforms SOTA iteration-free models. DFT validation further confirms that its predictions lie close to true ground-state energies and substantially reduce the number of ionic steps required for full optimization in DFT calculations.

\section{Acknowledgments}
This research was supported by the National Natural Science Foundation of China (Grant No. 62506143), the Natural Science Foundation of Guangdong Province (Grant No. 2025A1515011487), Ministry of Education, Singapore, Tier 1 (Grant No. A-8001194-00-00), and Tier 2  (Grant No. A-8001872-00-00).

\setcounter{secnumdepth}{2}          
\renewcommand\thesubsection{\Alph{subsection}}  
\setcounter{subsection}{0}

\appendix           

\section*{Appendix}

\subsection{Initialisation of Features}
The scalar feature $\bm{x}_i^{(0)}$ is initialized based on the atomic number, $E(z_i) \in \mathbb{R}^F$, where $z_i$ is the atomic number and $E$ is an embedding layer. The vector feature is initially set to zero, $\vec{\bm{x}}_i^{(0)} = \vec{\bm{0}} \in \mathbb{R}^{F \times 3}$. The atomic coordinates, $\vec{\bm{r}}_i^{(0)}$, are initialized to the coordinates of the initial unrelaxed structure. The lattice node features \(\bm{y}_c^{(0)}\) and \(\vec{\bm{y}}_c^{(0)}\) are initialized analogously to atom features: \(\bm{y}_c^{(0)} = E_{\text{lat}}(c)\), and \(\vec{\bm{y}}_c^{(0)} = \vec{\bm{0}} \in \mathbb{R}^{F \times 3}\), where \(E_{\text{lat}}\) is a learnable embedding for each lattice direction. The lattice coordinates, $\vec{\bm{l}}_c^{(0)}$, are initialized to the lattice vectors of the initial unrelaxed structure. 

\begin{table*}[]
\centering
\caption{Summary of the four benchmark datasets used in our experiments.}
\begin{tabular}{lllll}
\hline
Dataset                & \multicolumn{1}{c}{Dimensionality} & \multicolumn{1}{c}{Sample size} & \multicolumn{1}{c}{Elements covered} & \multicolumn{1}{c}{Train : Val : Test} \\ \hline
Materials Project (MP) & 3D                                 & 62 724                          & 89                                   & 8 : 1 : 1                              \\
X-Mn-O                 & 3D                                 & 28 579                          & 6                                    & 8 : 1 : 1                              \\
C2DB                   & 2D                                 & 11 581                          & 62                                   & 6 : 2 : 2                              \\
Layered vdW Crystals   & 2D                                 & 57                              & 29                                   & 8 : 1 : 1                              \\ \hline
\end{tabular}
\label{tbl:datasets}
\end{table*}

\subsection{Datasets}
We evaluate E$^{3}$Relax on four benchmark datasets: MP \cite{jain2013commentary}, X–Mn–O \cite{kim2020generative, kim2023structure}, C2DB \cite{gjerding2021recent, lyngby2022data}, and layered vdW crystals \cite{yang2024scalable}. A summary of these datasets is provided in Table~\ref{tbl:datasets}. 

The MP database spans 89 elements and contains 187,687 structure snapshots from 62,783 compounds, each sampled at initial, intermediate, and fully relaxed stages of DFT optimization. After removing entries missing either the initial or final structure, 62,724 usable pairs remain, which we split 8:1:1 into training, validation, and test sets. The hypothetical substitution dataset (X-Mn-O), derived from MP, focuses on ternary oxides for photoanode research. It covers four ‘X’ elements (Mg, Ca, Ba, Sr) substituted into prototype Mn–O frameworks, yielding 28,579 unrelaxed–relaxed pairs. These are partitioned 8:1:1 for training, validation, and testing. The C2DB dataset comprises 11,581 unrelaxed–relaxed 2D structure pairs spanning 62 elements, which we split into training/validation/test sets in a 6:2:2 ratio. The layered‐vdW crystal dataset contains 57 2D van der Waals–bonded structures across 29 elements, divided into training, validation, and test subsets in an 8:1:1 ratio.

\begin{figure}[!htb]
  \centering
  \includegraphics[width=1\columnwidth]{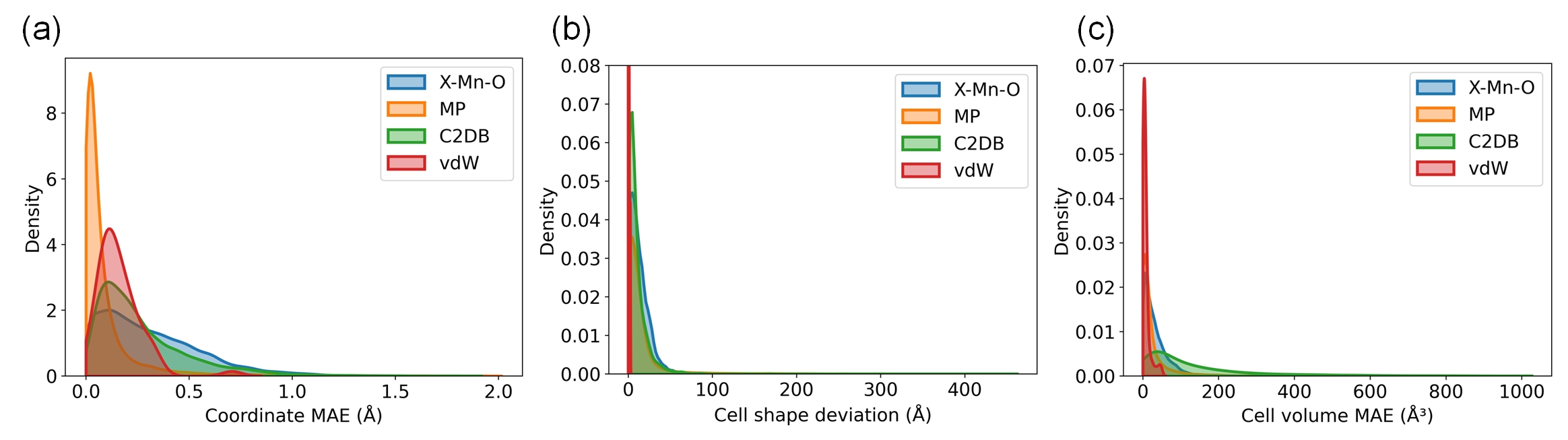}
  \caption{The deviation distribution of the four datasets: (a) coordinate MAE ($\rm \AA$), (b) cell shape deviation ($\rm \AA$), and (c) cell volume MAE ($\rm \AA^3$) between initial and final structures.}
  \label{fgr:deviation_distribution}
\end{figure}

Figure \ref{fgr:deviation_distribution}(a)–(c) compares the distributions of atomic coordinate MAE, cell shape deviation, and cell volume MAE between the unrelaxed and DFT‐relaxed structures for the four datasets. Three distinct patterns emerge. The MP dataset exhibits narrow distributions in all three metrics, indicating that most unrelaxed structures are already very close to their relaxed counterparts. In contrast, the X–Mn–O and C2DB datasets display long tails, reflecting substantial atomic displacements and lattice deformations during structural optimizations. The layered vdW crystals fall in between: they show modest coordinate MAEs but tightly clustered cell‐shape and volume errors near zero, consistent with the minimal lattice adjustments expected from weak van der Waals interlayer interactions.

\subsection{Baselines}
We compare E$^{3}$Relax against six SOTA iteration-free ML models: DeepRelax \cite{yang2024scalable}, PAINN \cite{schutt2021equivariant}, EGNN \cite{satorras2021n}, EquiformerV2 \cite{equiformer_v2}, HEGNN \cite{cen2024high}, and GotenNet \cite{aykent2025gotennet}.

\begin{itemize}
    \item \textbf{DeepRelax} first predicts interatomic distances, atomic displacements, and lattice parameters, then solves a Euclidean‐distance‐geometry optimization to reconstruct Cartesian coordinates of relaxed structures. To predict relaxed lattice vectors, it pools the final GNN node embeddings into a single global vector, concatenates this with a linear embedding of the initial lattice vectors, passes the result through a small feed‐forward network that outputs nine values, and reshapes them into a $3 \times 3$ matrix.
    \item \textbf{PAINN} is an equivariant graph neural network originally developed for interatomic potential and force prediction. Although not designed for iteration‐free optimization, we adapt its force‐prediction branch to output atomic displacements, and we use the same pooling‐and‐feedforward strategy as DeepRelax to predict the relaxed lattice.
    \item \textbf{EGNN} is a lightweight SE(3)‐equivariant architecture that updates atomic coordinates at each layer, enabling direct prediction of atomic coordinates of relaxed structures. We employ the identical lattice‐prediction pipeline used by DeepRelax.
    \item \textbf{EquiformerV2} is an equivariant Transformer that incorporates efficient eSCN convolutions, attention re‐normalization, separable S$^2$ activations, and separable layer normalization to model higher‐order geometric interactions. We adapt its force‐prediction head to output atomic displacements and apply the DeepRelax lattice‐prediction procedure.
    \item \textbf{HEGNN} generalizes EGNN by adding high-degree steerable representations through efficient inner-product scalarization, preserving equivariance while improving expressivity.
    \item \textbf{GotenNet} rethinks efficient 3D equivariant GNNs by combining geometric tensor attention and hierarchical refinement—achieving high expressiveness without the computational burden of Clebsch–Gordan tensor products.
\end{itemize}

\begin{figure}[!ht]
  \centering
  \includegraphics[width=1\columnwidth]{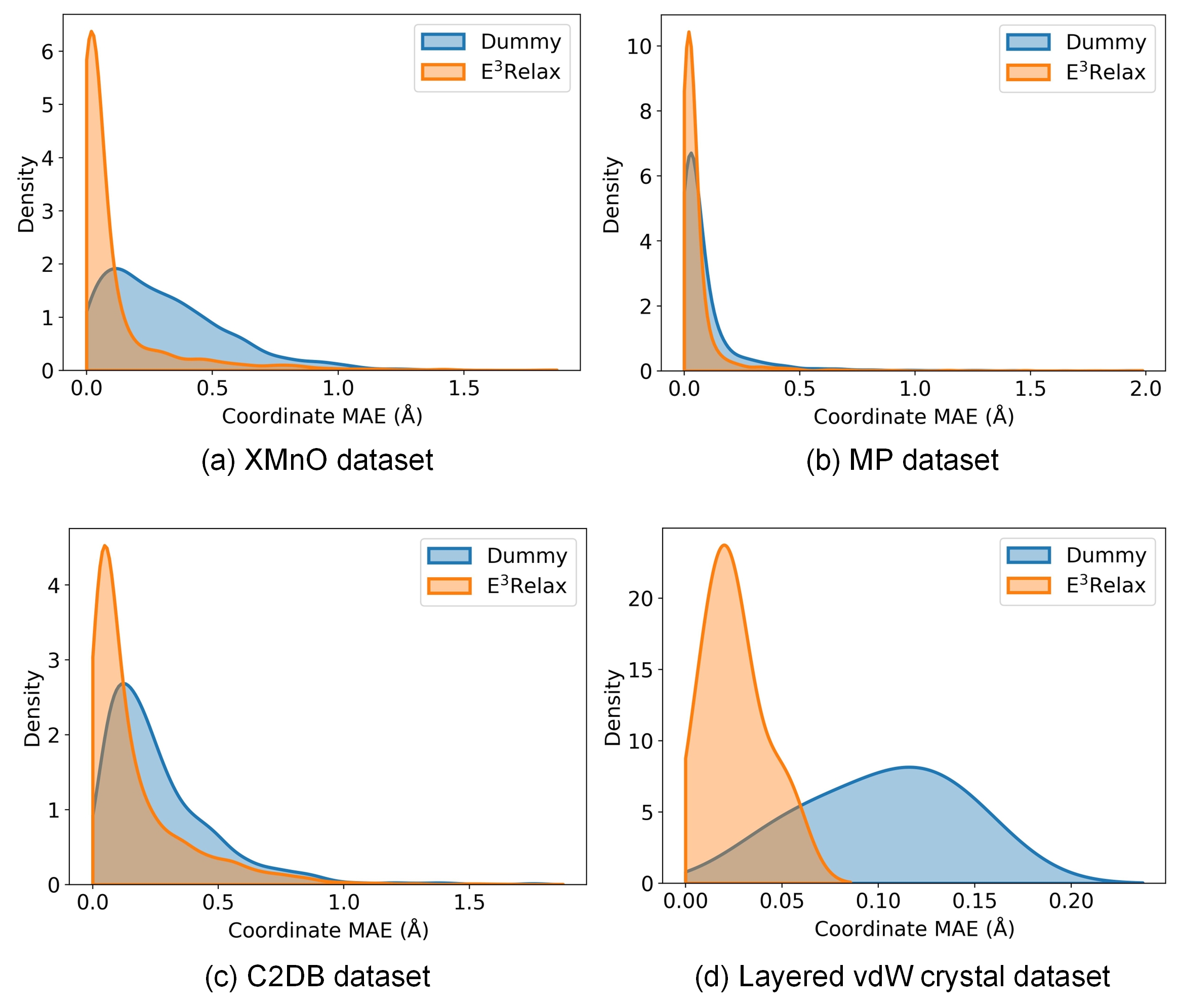}
  \caption{Distribution for coordinates MAE ($\rm \AA$) for structures predicted by the Dummy model and E$^{3}$Relax across various datasets. (a) X-Mn-O, (b) MP, (c) C2DB, and (d) layered vdW crystals.}
  \label{fgr:pred_distribution}
\end{figure}

\subsection{Performance Indicators}
\label{sec:metrics}
\subsubsection{MAE of Coordinate}
The MAE of coordinates quantifies the positional discrepancies between ML-predicted and DFT-relaxed structures. It is defined as follows:
\begin{equation}
\Delta_{\rm coord} = \frac{1}{3N} \sum_{v_i \in \mathcal{V}^a}\lvert \vec{\bm{r}}_i^{(T)} - \tilde{\bm{r}}_i \rvert  
\end{equation}
where $N$ is the total number of nodes in graph $\mathcal{G}$, $T$ is the total number of layers, and $\vec{\bm{r}}_i^{(T)}$ and $\tilde{\bm{r}}_i$ represent the predicted and ground truth Cartesian coordinates, respectively.

\subsubsection{Cell Shape Deviation}
We measure cell shape deviation by the Frobenius norm of the difference between the ML-predicted and DFT‐relaxed lattice metric tensors. Given the lattice matrix $\bm{L} = [\vec{\bm{l}}_1, \vec{\bm{l}}_2, \vec{\bm{l}}_3]$, the lattice metric tensor is $\bm{G} = \bm{L}^T \bm{L}$. The cell shape deviation is then
\begin{equation}
\Delta_{\rm shape} = \sqrt{\sum_{i=1}^{3} \sum_{j=1}^{3} (\bm{G}_{\text{pred}}^{ij} - \bm{G}_{\text{true}}^{ij})^2}
\end{equation}
where $\bm{G}_{\text{pred}}^{ij}$ and $\bm{G}_{\text{true}}^{ij}$ represent the components of the predicted and truth lattice metric tensors, respectively.

\subsubsection{MAE of Cell Volume}
The cell volume error measures the model’s ability to preserve the overall scale of the unit cell. It is defined as the absolute difference between the predicted and reference volumes:
\begin{equation} 
\Delta_{\rm volume}=\Bigl\lvert \vert \vec{\bm{l}}_1^{(T)} \cdot (\vec{\bm{l}}_2^{(T)} \times \vec{\bm{l}}_3^{(T)}) \vert - \vert \tilde{\bm{l}}_1 \cdot (\tilde{\bm{l}}_2 \times \tilde{\bm{l}}_3) \vert \Bigr\rvert
\end{equation}
where $\vec{\bm{l}}_1^{(T)}$, $\vec{\bm{l}}_2^{(T)}$, and $\vec{\bm{l}}_3^{(T)}$ are the predicted lattice vectors, and $\tilde{\bm{l}}_1$, $\tilde{\bm{l}}_2$, and $\tilde{\bm{l}}_3$ are their corresponding DFT-relaxed reference lattice vectors.

\subsection{Distribution of Predicted Structures}
Figure \ref{fgr:pred_distribution} compares the coordinate MAE distributions obtained with the Dummy model (blue) and E$^{3}$Relax (orange) across the four datasets. In every case the E$^{3}$Relax curve is shifted markedly toward lower errors, confirming that the model recovers geometries much closer to the DFT-relaxed ground truth. The gain is most significant for the X–Mn–O, C2DB and layered-vdW sets, where large structural optimizations occur. For the MP dataset, however, the improvement over the Dummy model is less visually pronounced. This is attributed to the relatively minor structural differences between the unrelaxed and relaxed structures in the MP dataset, as shown in Figure \ref{fgr:deviation_distribution}.

\begin{figure}[htb]
  \centering
  \includegraphics[width=1\columnwidth]{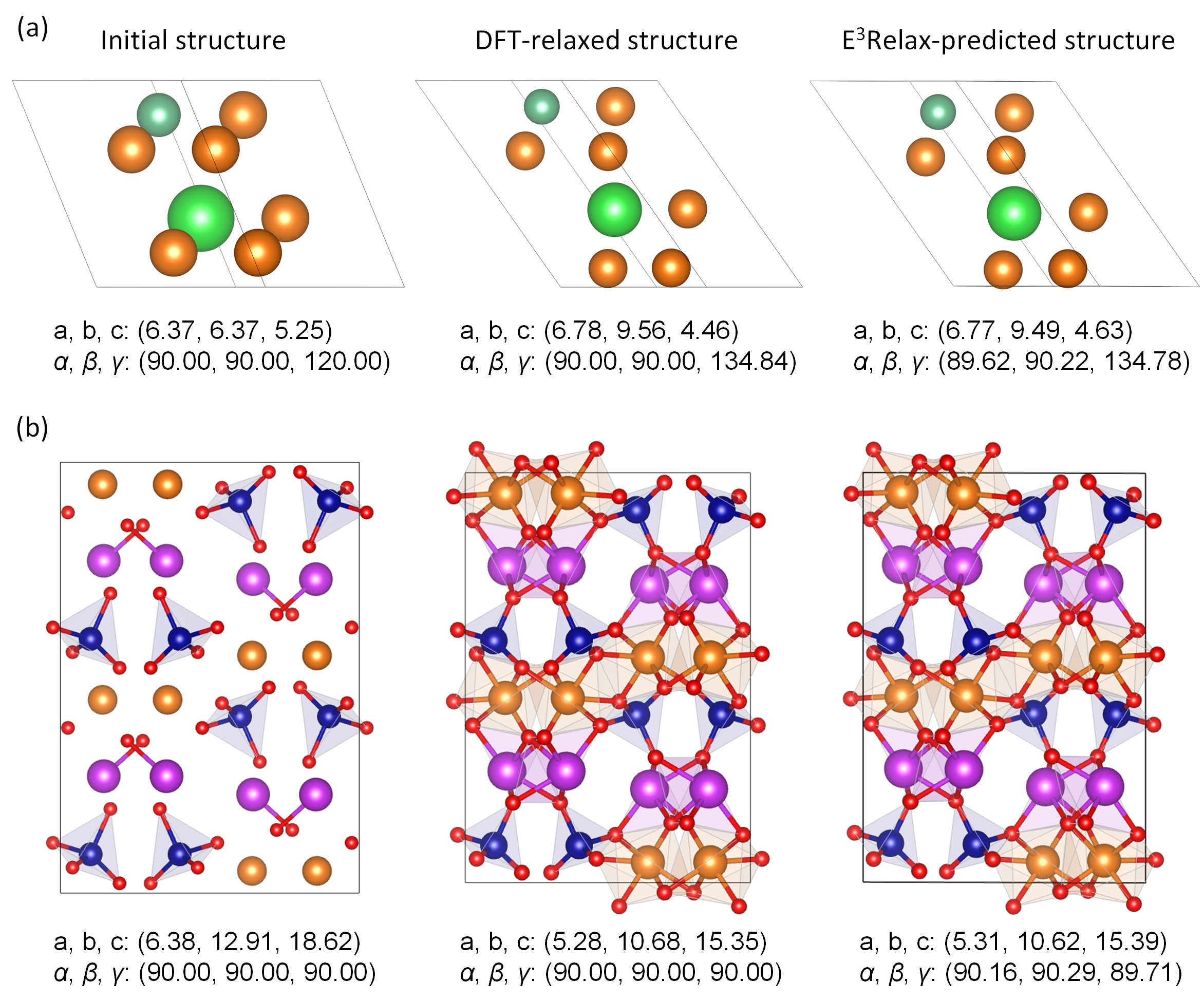}
  \caption{Two crystal structures from the MP database after optimization with E$^{3}$Relax: (a) $\rm BaMg_6Nb$ and (b) $\rm Bi_8Cr_8Mg_8O_{40}$. \(a\), \(b\), and \(c\) are lattice constants in angstroms (Å), and \(\alpha\), \(\beta\), and \(\gamma\) are angles in degrees ($^\circ$).}
  \label{fgr:mp}
\end{figure}

\begin{figure}[!htb]
  \centering
  \includegraphics[width=1\columnwidth]{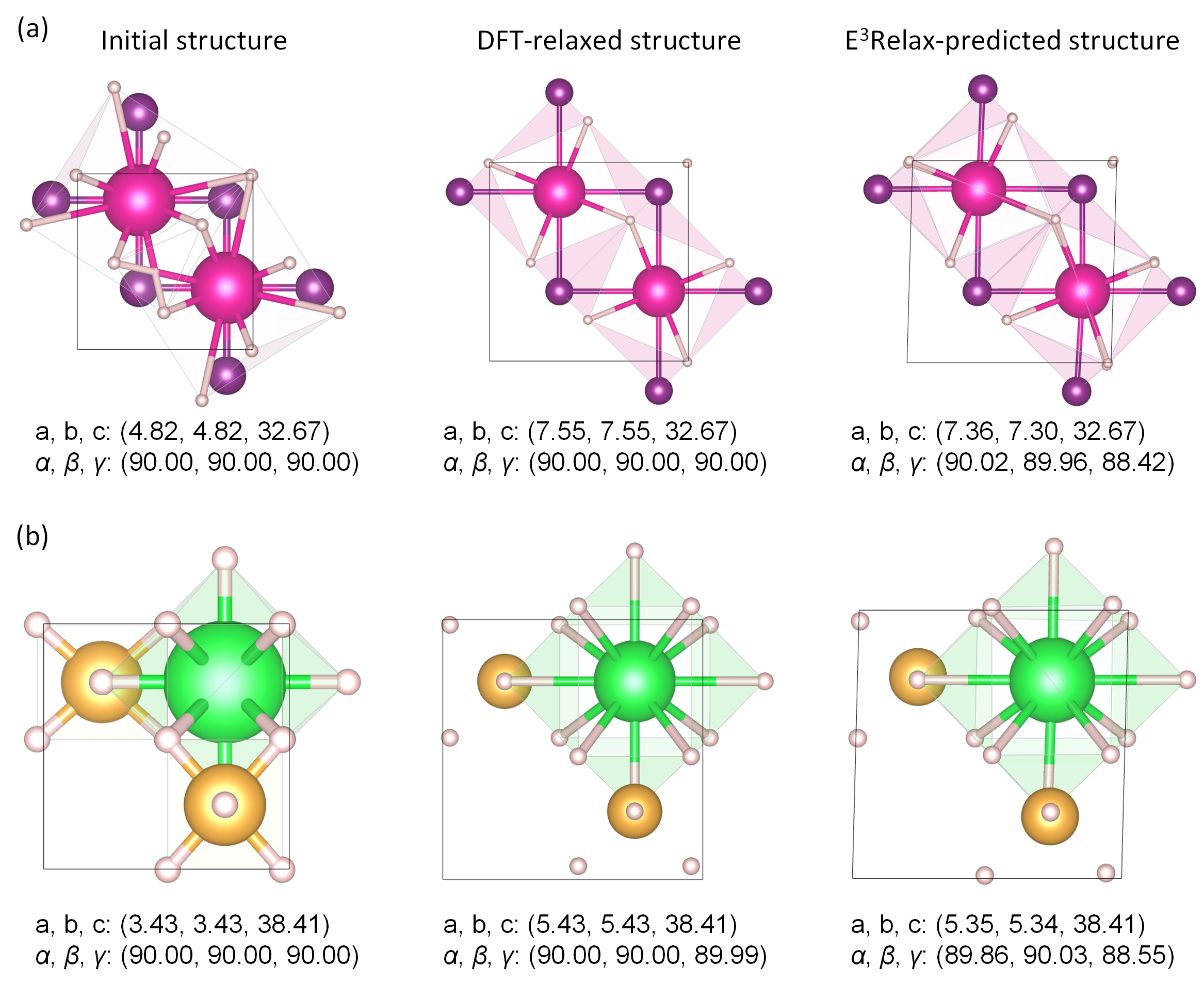}
  \caption{Visualization of two crystal structures from the C2DB dataset relaxed by E$^3$Relax: (a) $\rm H_8I_2Rb_2$  and (b) $\rm $$\rm H_{12}Au_2Sr$, where \(a\), \(b\), and \(c\) are lattice constants in angstroms (Å), and \(\alpha\), \(\beta\), and \(\gamma\) are angles in degrees ($^\circ$).}
  \label{fgr:c2db}
\end{figure}

\begin{figure}[!ht]
  \centering
  \includegraphics[width=1\columnwidth]{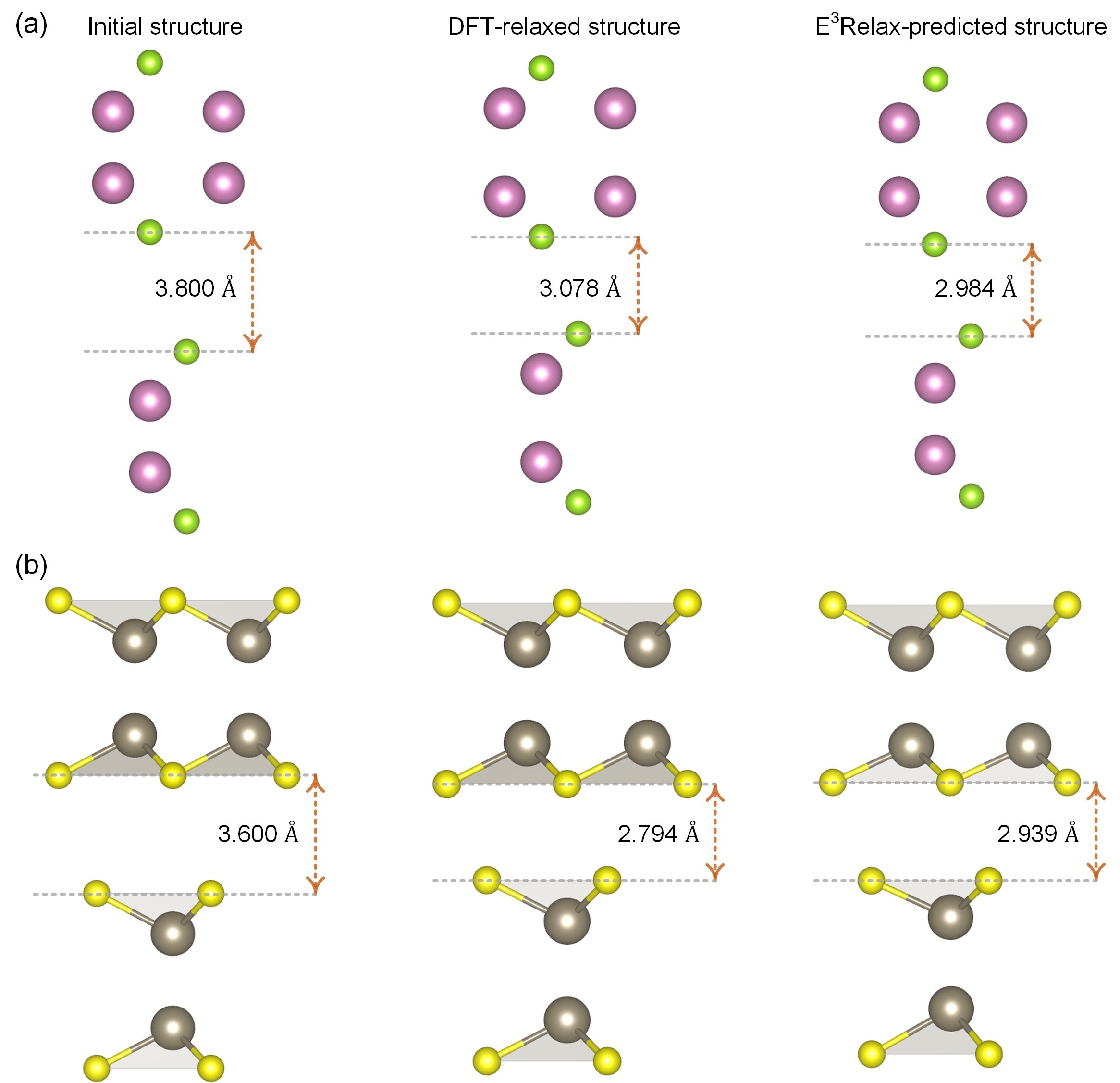}
  \caption{Visualization of two layered vdW crystal structures relaxed by E$^{3}$Relax: (a) $\rm In_4 Se_4$. (b) $\rm S_4 Tl_4$.}
  \label{fgr:vdW_vis}
\end{figure}

\subsection{Visualization of E$^3$Relax-predicted Structures}
To further qualitatively evaluate E$^{3}$Relax, we selected representative test‐set examples from MP, C2DB, and layered vdW crystals dataset and relaxed them using our model (Figures \ref{fgr:mp}–\ref{fgr:vdW_vis}). In these examples, the E$^{3}$Relax–predicted geometries closely match the corresponding DFT‐relaxed structures, particularly in terms of atomic arrangements.

\subsection{Hyperparameter Sensitivity Analysis}
\begin{figure}[!ht]
  \centering
  \includegraphics[width=1\columnwidth]{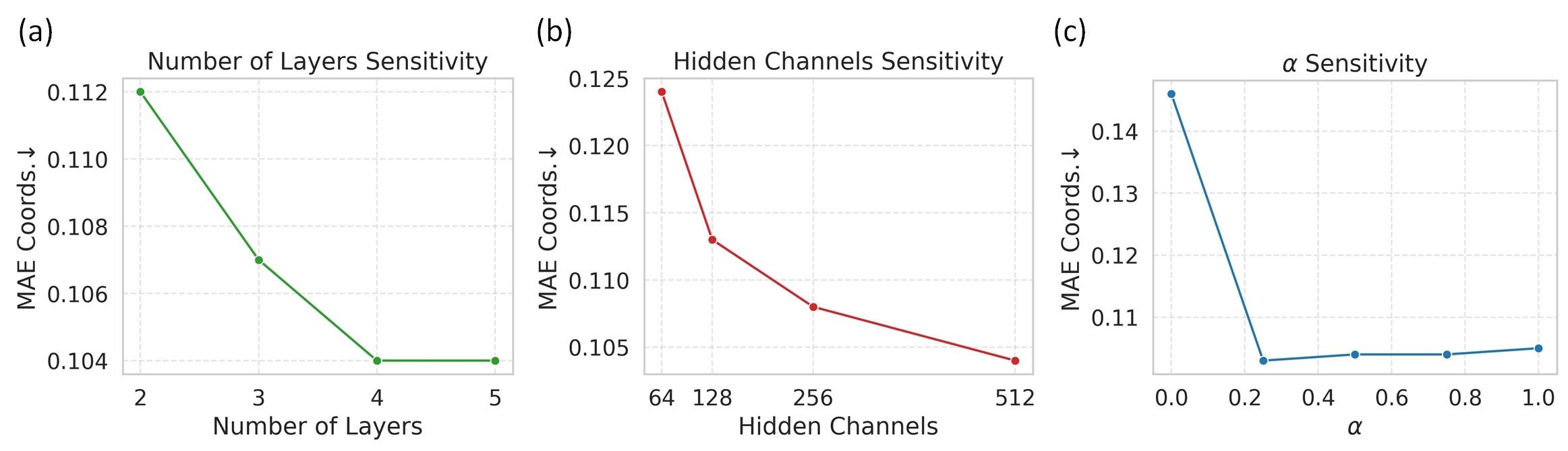}
  \caption{Sensitivity analysis of E$^{3}$Relax with respect to network depth ($T$), hidden dimension ($F$), and layer weights ($\alpha_t$).}
  \label{fgr:sensitivity}
\end{figure}

We further analyze the sensitivity of E$^{3}$Relax to key hyperparameters, including the number of layers ($T$), hidden dimension size ($F$), and layer weights ($\alpha_t$) (Figure~\ref{fgr:sensitivity}). Overall, E$^{3}$Relax demonstrates strong robustness to hyperparameter variations—its performance remains stable across a broad range of network depths, hidden dimensions, and layer-weight configurations.

\subsection{Training M3GNet and CHGNet}
In addition to iteration-free models, another popular approach for accelerating structural optimization is iteration-based methods. These methods train GNN-based surrogate models, such as M3GNet \cite{chen2022universal}, CHGNet \cite{deng2023chgnet}, and MACE \cite{batatia2022mace}, to approximate energy, forces, and stress. These predictions then drive an outer optimization loop, in which both atomic positions and lattice parameters are updated repeatedly until convergence.  

In this study, we selected two well-known and straightforward-to-implement iterative models, M3GNet \cite{chen2022universal} and CHGNet \cite{deng2023chgnet}, as baseline models. These models were implemented directly from the authors' public repositories using default settings. We follow the original M3GNet protocol \cite{chen2022universal} by extracting three snapshots per trajectory—the initial, an intermediate, and the final relaxed structures—yielding three structures per sample. The two models were then trained to predict energy, forces, and stress. Once trained, the models were used to relax crystal structures by iteratively guiding atoms toward lower-energy regions.

\subsection{DFT Parameters}
In this study, DFT calculations are conducted using the Vienna Ab initio Simulation Package (VASP) \cite{kresse1996efficient} and Atomic Simulation Environment (ASE) Python library \cite{larsen2017atomic}. These calculations employ the Generalized Gradient Approximation (GGA) with the Perdew-Burke-Ernzerhof (PBE) exchange-correlation functional \cite{perdew1996generalized}. For the XMnO dataset, we set the energy cut-off at 500 eV, the energy convergence criterion at $1.0 \times 10^{-5}$ eV, and use a K-point mesh of $6 \times 6 \times 6$. Gaussian smearing with a width of 0.05 eV is used for the electronic occupations. For the C2DB dataset, the energy convergence criterion is set to $1.0 \times 10^{-4}$ eV, and the K-point mesh is set to $6 \times 6 \times 1$. For layered vdW crystals, DFT calculations were conducted with van der Waals corrections using the DFT-D3 Grimme method. For $\rm MoS_2$ structure with point defects, optimization was performed until interatomic forces were below $0.05 \ \rm{eV/\AA}$. Spin polarization was included, following the approaches of previous studies \cite{huang2023unveiling, kazeev2023sparse}.

\bibliography{aaai2026}

\end{document}